\documentclass[aps,showpacs]{revtex4}

\usepackage{amsmath}
\usepackage{amsfonts}
\usepackage{amssymb}
\usepackage{color}

\def\bc{\begin{center}}
\def\nno{\nonumber}
\def\ec{\end{center}}
\def\be{\begin{eqnarray}}
\def\ee{\end{eqnarray}}

\newcommand{\ul}{\underline}

\newcommand{\omits}[1]{}

\definecolor{dyellow}{rgb}{1.,0.8,.0}
\definecolor{myblue}{rgb}{.1,.1,.7}
\definecolor{dcyan}{rgb}{.0,.6,.6}
\definecolor{dmagenta}{rgb}{0.6,0.0,0.6}
\definecolor{brown}{rgb}{0.6,0.2,0.}
\definecolor{darkblue}{rgb}{.0,.0,0.5}
\definecolor{darkred}{rgb}{0.75,0.0,0.0}
\definecolor{orange}{rgb}{1.,.6,.0}
\definecolor{dorange}{rgb}{0.8,.4,.0}
\definecolor{darkgreen}{rgb}{0.0,0.6,0.0}
\definecolor{purple}{rgb}{.4,.0,.4}

\def\blue{\color{blue}}

\def\La{\Lambda}
\def\Si{\Sigma}

\def\dl{\delta}
\def\eps{\epsilon}

\def\si{\sigma}

\def\d#1#2{\frac{\displaystyle #1}{\displaystyle #2}}

\newcommand{\dS}{$d{\cal S}$}
\newcommand{\AdS}{${\cal A}d\cal S$}
\newcommand{\BdS}{${\cal B}d{\cal S}$}

\newcommand{\HlsM}{${H}_l \subset M^{1,4}$}
\newcommand{\HrsM}{${H}_R \subset M^{1,4}$}

\newcommand{\GR}{general relativity}
\newcommand{\SR}{relativity}
\newcommand{\IWR}{inverse Wick rotation}

\newcommand{\PoR}{principle of relativity}
\newcommand{\CP}{cosmological principle}

\newcommand{\OoI}{${\cal O}o{\cal I}$}
\newcommand{\EoM}{${\cal E}o{\cal M}$}
\newcommand{\OoIm}{${\cal O}o{\cal I}m$}

\newcommand{\Mink}{${\cal M}ink$}

\newcommand\btd{\raise 2pt
\hbox{$\hat\bigtriangledown$}\hskip 1.5pt}
\newcommand\bt{\raise 2pt
\hbox{$\bigtriangledown$}\hskip 1.5pt}

\newcommand{\FLT}{$\cal FLT$}

\newcommand{\E}{$\cal E$}
\newcommand{\Lo}{$\cal L$}

\def\PRD{{\it Phys. Rev.}~{\bf D}}

\def\PLA{{\it Phys. Lett.}~{\bf A}}

\omits{

\newcommand{\omits}[1]{}

\def\bc{\begin{center}}
\def\nno{\nonumber}
\def\ec{\end{center}}
\def\be{\begin{eqnarray}}
\def\ee{\end{eqnarray}}
\newcommand{\ul}{\underline}

\def\La{\Lambda}
\def\Si{\Sigma}

\def\dl{\delta}
\def\eps{\epsilon}

\def\si{\sigma}

\def\d#1#2{\frac{\displaystyle #1}{\displaystyle #2}}

\newcommand\btd{\raise 2pt
\hbox{$\hat\bigtriangledown$}\hskip 1.5pt}
\newcommand\bt{\raise 2pt
\hbox{$\bigtriangledown$}\hskip 1.5pt}

}

\begin{document}

\title{ The Beltrami Model of De Sitter Space:\\[1mm]
From Snyder's quantized space-time to de Sitter invariant
relativity\footnote{Invited talk given at `International workshop
on
 noncommutative geometry and physics', Beijing, Nov. 7-10, 2005. To appear in the proceedings.}} 

\author{Han-Ying Guo\thanks{Email: \tt hyguo@itp.ac.cn}}

\affiliation{%
Institute of Theoretical Physics, Chinese Academy of Sciences,
Beijing 100080, China.}

\begin{abstract}
In terms of the Belrami model of de Sitter space we show that there
is an interchangeable relation between Snyder's quantized space-time
model in \dS-space of momenta at the Planck length $\ell_P=(G\hbar
c^{-3})^{1/2}$ and the \dS-invariant special relativity in
\dS-spacetime of radius $R\simeq(3\Lambda^{-1})^{1/2}$, which is
 another fundamental length related to the
cosmological constant. Here, the cosmological constant $\Lambda$ is
regarded as a fundamental constant together with the speed of light
$c$, Newton constant $G$ and Planck constant $\hbar$. Furthermore,
the physics at  two fundamental scales of length,  the \dS-radius
$R$ and the Planck length $\ell_P$, should be dual to each other and
linked via the gravity with local \dS-invariance characterized by a
dimensionless coupling constant $g= \sqrt{3} \ell_P/R\simeq( G\hbar
c^{-3}\Lambda)^{1/2}\sim 10^{-61}$.
\end{abstract}

\pacs{04.90.+e, 
04.50+h, 
03.30.+p, 
02.40.Dr, 
02.40.Gh  
} \maketitle

\tableofcontents

\section{Introduction}

Long time ago,  Snyder proposed a quantized space-time
model\cite{Snyder}, in which 
the space-time coordinates are no longer ordinary real numbers
rather some
non-commutative operators. 
Snyder started with a projective geometry approach to the de
Sitter (\dS)-space of momenta with  
a large energy-momentum scale $a$ near or at the Planck scale.
Denoting  the energy and momentum by means of the inhomogeneous
projective coordinates,
  the space-time
coordinates' operators $\hat x_j$ are defined by 4-`translation'
Killing vectors of $so(1,4)$-algebra\omits{together with others
form an $so(1,3)$, the homogeneous Lorentz algebra, in \dS-space
of momenta}. Thus, $\hat x_j$ are noncommutative.

Recently, in order to explain the Greisen-Zatsepin-Kuz'min 
effects \cite{GZK} the 
`doubly spacial relativity' (DSR) models have been proposed
\cite{DSR}. In DSR, there is also a large  energy-momentum scale $\kappa$ near the Planck scale 
 in addition to  the speed of light $c$. It is found that there is a close relation
between Snyder's model and DSR. In fact, DSR can be regarded as generalization of Snyder's model \cite{DSRdS}  and most DSR models with 
$\kappa$-Poincar\'e algebra can be realized geometrically by means
of particular coordinate systems on \dS, Minkowski (\Mink) and
 anti-\dS\ (\AdS)-space of momenta \cite{DSRdS} other than the
inhomogeneous projective coordinates used by Snyder for the
\dS-space of momenta, or its counterpart in the \AdS-space of
momenta. Thus, in this point of view, there is a kind of
coordinate transformations from Snyder's model to a kind of DSR in
\dS/\AdS-space of momenta and vise versa, respectively.


For  the \dS-space the projective geometry model can be completely
substituted by the Beltrami-like model (the Beltrami model for
short) \cite{beltrami, klein1871} (see also \cite{R}). It is very
important that in either Beltrami coordinates or inhomogeneous
projective ones\omits{, which are ignored for long time in general
relativity, might be partially due to the coordinate-independence
hypothesis.} the timelike and null geodesics in \dS-space are in
linear forms and all these properties are invariant under the
fractional linear transformations with common
denominator (\FLT s) 
of the \dS-group $SO(1,4)$. Thus, the particles and light signals
move along these timelike and null geodesics, respectively, are
all with {\it constant} coordinate velocities. Therefore, these
particles and signals look like in free motion  of inertia in a
space without gravity. Namely, these motions should be regarded as
a new kind of inertial motions in \dS-space. For the definiteness,
we focus on the Beltrami-coordinates and name the \dS-space with
such coordinates as the Beltrami-\dS (\BdS)-space.\omits{Namely,
among 
different coordinate systems of \dS-spaces, there is a kind of
important ones  in analog with the \Mink-coordinates in the
\Mink-space. It is the Beltrami coordinates for the \dS-space,
called the Beltrami-de Sitter (\BdS) space for short, and it is
precisely the Beltrami model
 of a \dS-space. \omits{hyperboloid ${\cal H}_R$ embedded in a 5-d \Mink-space,  \BdS\ $\backsimeq$ \HrsM. }By means of 
 the Beltrami coordinate systems the \BdS-space can be covered
patch by patch, in which particles and light signals move along
the timelike and null geodesics, respectively, with {\it constant}
coordinate velocities. And all these properties are invariant
under the fractional  linear transformations with common
denominator (\FLT s) 
of the \dS-group $SO(1,4)$, among the Beltrami systems of the
\dS-space. Therefore, these particles and signals look like in
free motion  of inertia in a space without gravity. Thus, the
Beltrami coordinates and observers at these systems should be of
inertia.}

Historically, de Sitter \cite{dS17}   first  used the Beltrami
coordinates  for the spacetime of constant curvature in 1917, the
same year he found his solution, in the course of the debate with
Einstein on `relative inertia'. A few years later, Pauli, in his
famous book \cite{Pauli20}, noticed the Beltrami metric with
Euclid signature, the Beltrami model of Riemann sphere, and asked
for its physical application.  \omits{ If those important issues
would be noticed physically, neither de Sitter and Einstein
noticed this important property nor Snyder and Pauli transferred
the model to \dS-spacetime; otherwise the \dS-invariant inertial
law should be discovered long ago.}However, there had been no
study on the key issues for many years until early 1970s. In 1970,
Lu \cite{Lu} first emphasized these issues of \dS/\AdS-space and
asked `why we must use the Minkowski-metric?' A few years later,
with his collaborators he began to study the special relativity in
\dS/\AdS-space \cite{LZG}. Recently, promoted by the observations
on the dark universe, the study has been made further
\cite{Lu05}-\cite{T}.

In fact, there should be three kinds of relativity in the
\Mink/\dS/\AdS-space with Poincar\'e, \dS\ or \AdS\ invariance,
respectively, on almost equal footing. The existence of three kinds
of special relativity can be understood as the physical counterparts
of the three kinds of geometry, the Euclid(-flat), Riemann-spherical
and Lobachevsky-hyperbolic geometry, on almost equal footing except
the fifth axiom on parallels. In these geometries of constant
curvature in 4 dimensions, there are Descartes or Beltrami
coordinate systems as a kind of special coordinate systems for
Euclidean or non-Euclidean geometry, respectively. In these systems,
the points, straight-lines and metric are invariant or mutually
transformed under linear transformations of $ISO(4)$ for
Euclid-space or the \omits{fractional linear ones with common
denominators }\FLT s of $SO(5),\,SO(1,4)$ for Riemann-sphere and
Lobachevsky-hyperboloid, respectively. Beltrami \cite{beltrami}
introduced such coordinates for description of Lobachevsky-plane
first and completed by Klein \cite{klein1871}. Under
 an inverse Wick rotation \cite{IWR}, these constant curvature spaces  become
$ISO(1,3), SO(1,4)$ and $SO(2,3)$-invariant \Mink, \dS\ and
\AdS-spacetime, respectively,
 with events, straight
world-liners and Minkowski, or Beltrami-metric of physical
signature, say, {$(+,-,-,-)$}. Thus, Klein's  invariance program
for geometry under transformation groups \cite{klein} should imply
the principle of relativity in those maximally
symmetric spacetimes.

Einstein's special relativity 
is set up on the \Mink-spacetime as the counterpart of
Euclid-space, on the \dS/\AdS-spacetime as the non-Euclidian
counterpart of 4-d Riemann-sphere and Lobachevski-hyperboloid
there is just \dS/\AdS-invariant special \SR, respectively,
based on the principle of relativity and the postulate on
universal constants of $c$ and $R$ as the curvature radius of the
\dS/\AdS-spacetime. \omits{The principle require: {\it the
physical laws without gravity are invariant under the group
transformations among  the inertial-systems}. The postulate
states: {\it there are two universal constants,  the speed of
light $c$ and curvature-radius $R$}.}

Thus, there is an inertial law for free particles and light
signals in \dS/AdS-spacetime. It is similar to 
the inertial law in either Newton mechanics or Einstein's special
relativity. Accordingly, as a momentum-`picture' in quantum
mechanics, there should be an inertial-like law for the `group
velocity' of some `wave packets' of free particles in Snyder's
model on \dS/\AdS-space of momenta. 

It is important and interesting that in terms of the Beltrami
model of \dS-space, 
there is
an interchangeable dual relation between Snyder's quantized
space-time model \cite{Snyder} as a simplest and earliest DSR
\cite{DSR,DSRdS}
 and \dS-invariant  relativity \cite{Lu, LZG, Lu05, BdS, IWR, BdS2,  NH, T}. \omits{as was briefly mentioned in \cite{BdS}.
 : `we may directly regard
${\cal SR}_{c,R}$ as the counterpart of DSR in its momentum space
without orientation problem as long as $R$ is taken as the
observer-independent large-momentum scale'. }In addition, from
Snyder's constant $a$ (similar to $\kappa$ in DSR) or the Planck
length $\ell_P$ and the cosmological constant $\Lambda$, it
follows a dimensionless constant $g$
 \be\label{g}%
 \iota:=\kappa^2/R^2 \to g^2:=3{\ell_P}^2R^{-2}\simeq G\hbar c^{-3}\Lambda\simeq 10^{-122}.
 \ee%
This dimensionless constant 
$g$ has been appeared in a kind of simple models of
\dS/\AdS-gravity \cite{dSG, T77, QG}), to characterize the
self-interaction of gravity with local \dS/\AdS-invariance,
respectively.

 Thus, we may propose such a conjectured Planck scale-cosmological constant duality: 
The cosmological constant should be a fundamental constant together
with the Planck constant $\hbar$, Newton gravitational constant $G$
and $c$. The physics at such two scales should be dual to each other
and linked via gravity of local \dS-invariance characterized by a
dimensionless coupling constant $g$. 

 Thus, there no longer exist 
 the puzzle on $\La$ as the `vacuum energy'. It should transfer
 to: What is the origin of  the dimensionless
 constant $g$?
 Is $g$  calculable?

 In this talk, we first briefly recall Snyder's model in
 \dS-space of momenta in section 2. Then we  introduce  the key issues of the Beltrami
 model of a Riemann sphere and that of a \dS-space via an inverse Wick rotation in section 3.
  We briefly introduce the \dS-invariant relativity  in section 4. We  show the
 interchangeable relation between Snyder's model and \dS-invariant  relativity and
 propose the Planck scale-cosmological constant
 duality in section 5. We end with some concluding remarks.

%
%

\section{Snyder's Quantized Space-time and DSR }

Snyder considered a homogenous quadratic form%
 \be\label{dsp}%
-\eta^2=\eta^2_0-\eta^2_1-\eta^2_2-\eta^2_3-\eta^2_4:=\eta^{AB}\eta_A\eta_B<0,\\\nno
(\eta^{AB})_{A, B=0, \cdots, 4}=(\eta_{AB})=diag(1,-1,-1,-1,-1),
\ee%
where $\eta_A$ may be regarded as the homogeneous (projective)
coordinates of a real 4-d space of constant curvature, a
\dS-space. According to Snyder, this was inspired by Pauli.

Then Snyder defined the energy-momentum 
with  a natural unit of length $a$%
\be\label{BdSp}%
p_0=\frac{ 1}{a} \frac{\eta_0}{\eta_4},\quad
p_\alpha:=\frac{1}{a}
\frac{\eta_\alpha}{\eta_4},\quad\hbar=1,\quad
\alpha=1, 2, 3.
\ee%
Quantum mechanically, in this `momentum picture' the operators for
the time coordinate and the space coordinates denoted as
 $ \hat t, \hat x_\alpha$ should be given by:
\be\label{xt}%
\hat x_\alpha &:=&i[\frac{\partial}{\partial p_\alpha}+a^2p_\alpha
p_j\frac{\partial}{\partial p_j}],\quad j=0,\cdots,3,\\\nno%
\hat x_0&:=&i[\frac{\partial}{\partial p_0}-a^2p_0
p_j\frac{\partial}{\partial p_j}], ~~x_0=c\hat t. 
\ee%
They are no longer commutative rather noncommutative `quantized'
operators. According to Snyder, they form an $so(1.4)$ algebra (in
what follows all hats \^\, are omitted) together with the `boost'
$M_\alpha$ and `3-angular momentum' $L_\alpha$ in the space of
momenta:%
 \be\label{so14m}%
[x_\alpha, x_\beta]={ia^2}L_\gamma,&&
[ t, x_\alpha ]={ia^2}M_\alpha,\\\nno%
[L_\alpha, L_\beta]=\epsilon_{\alpha\beta\gamma}L_\gamma,&&
[M_\alpha, M_\beta]=\epsilon_{\alpha\beta\gamma}M_\gamma;~~
etc.
\ee%
 Here $L_\alpha=x_\beta p_\gamma-x_\gamma p_\beta,~
M_\alpha=x_\alpha p_0+x_0p_\alpha.$

As was mentioned earlier, Snyder's model is a special case of the
DSR \cite{DSR, DSRdS}.
 In DSR \cite{DSR}, there is a large  energy-momentum scale $\kappa$ near the Planck scale 
 in addition to
 $c$ for all observers. This scale is just similar to the constant $a$ in
 Snyder's model. \omits{

 Motivation: GZK via modification of dispersion relation and energy conservation in \SR, etc.\\

 DSR (with 
$\kappa$-Poincar\'e algebra) can be realized via  particular
coordinate systems on \dS/$M$/\AdS-space of momenta \cite{DSRdS,
BdS}.}

Some remarks should be made in order. 
The energy-momentum $p_j$ are inhomogeneous (projective)
coordinates and one coordinate patch is not enough to cover the
\dS-space of momenta, which can be covered patch-by-patch. Since
the 4-d projective space $RP^4$ is not orientable, in order to
preserve the orientation, the antipodal identification should not
be taken. Actually, this model
can also be realized in terms of 
the Beltrami model of \dS-space of momenta as was mentioned
earlier and will be shown later. In this model, the operators of
$\hat x_j$ are 4-Killing vectors of the \dS-algebra. Other
operators $L_\alpha, M_\beta$ are just rest 6-Killing vectors
forming a homogeneous Lorentz algebra $so(1,3)$ in the momentum
space. Thus, similar to Snyder's model of \dS-space of momenta, a
model of
  \AdS-space of momenta as anti-Snyder's quantized space-time cam also be constructed. %
And the  relation between the Snyder-like quantized space-time models and DSR with 
$\kappa$-Poincar\'e algebra may be described as
the Beltrami coordinates for these models and other  
particular coordinate systems on 4-d \dS/\AdS-space of momenta for
other DSR models \cite{DSRdS}.

It is  important  that there are also other remarkable issues in the
Beltrami model of \dS/\AdS-spaces and we should consider  their
physical meaning. The most important issue is that in either the
Beltrami model or Snyder's projective geometry model, the geodesics
are all in linear forms in either Beltrami-coordinates or the
inhomogeneous projective coordinates, respectively. Namely, there
are {\it the straight `world'-lines} in \dS-space of momenta. In
fact, in Snyder's approach, these straight `world'-lines are just
the projective straightlines  of momenta. Is there any important
physical meaning for them?  In \dS-space of spacetime, their
counterparts are straight world-lines for the test particles and
light signals, should their motion be of \dS-invariant inertia? In
addition, there is a horizon and according to the standard approach
in \GR\ there should be some Hawking temperature and entropy
\cite{GH77} in some \dS-spacetimes. What about such kind of
properties of the horizon in the \dS-space of momenta? We may keep
these issues in mind for the moment.


\omits{}

%
%

\section{The Beltrami Model}

\subsection{The Beltarami model of  a Riemann sphere}


Let us
 focus on the Beltrami model of a 4-d Riemann sphere ${\cal S}^4$ with positive  constant
curvature, since its inverse
 Wick rotation is just  the Beltrami model of the
 \dS-space.  Similar issues appear in a 4-d Lobachevski space \Lo$^4$ with its inverse
 Wick rotation as  the Beltrami model of the
 \AdS-space.

The Riemann sphere ${\cal S}^4$  can be embedded
in a 5-d Euclid space \E$^5$
\be\label{4s}%
{\cal S}^4:~&&\delta_{AB}\xi^A \xi^B=l^2>0, \quad A, B=0, \cdots, 4,\\\label{5ds}%
&& ds_E^2=\delta_{AB}d\xi^A d\xi^B=d\xi {\cal
I}d\xi^t,\\\label{pbdry}%
\partial_P {\cal S}^4:&& \delta_{AB}\xi^A \xi^B=0, %
\ee%
where $\partial_P$ is the projective boundary. They are invariant under (linear) rotations of 
$
 SO(5)$:
\be\label{so5}%
 \xi~ \rightarrow ~\xi'=S~\xi, \quad S {\cal I} S^t={\cal
I}, ~~\forall ~S ~\in ~{SO(5)}.
\ee%
\omits{They are invariant under (linear) transformations 
of 5-d rotation group $SO(5)$.}

The Beltrami model describes an intrinsic geometry of ${\cal S}^4$
in the
Beltarmi-space 
${\cal B}_l$ 
with Beltrami coordinates: \omits{ i.e. the inhomogeneous
coordinates in projective geometry \cite{Rosenfeld, beltrami}
(hereafter, we call them after Beltrami
for short),}%
\be\label{Bcrd}%
x^i:=l\frac{\xi^i}{\xi^4},
\quad \xi^4\neq 0  ,\quad
 i=0,\cdots, 3.%
\ee%
To cover  ${\cal B}_l\sim$${\cal S}^4$, one Beltrami patch is not enough, but 
all properties of ${\cal S}^4$ should be well-defined in the
Beltrami model patch by patch \cite{BdS}. For simplicity, we
illustrate in one patch only.

The  metric (\ref{5ds}),  the sphere (\ref{4s}) and the boundary
(\ref{pbdry})
\omits{%
ds_E^2=\delta_{AB}d\xi^A d\xi^B %
\ee}%
 restricted
 on ${\cal B}_l$ become the Beltrami-metric, a domain condition and a boundary condition, respectively, as follows:
 \be\label{4Bds}%
 {\cal B}_l&& ds_E^2=\{\delta_{ij}\sigma_E^{-1}(x)-l^{-2}\sigma_E^{-2}(x)\delta_{il}x^l\delta_{jk}x^k\}
 dx^idx^j,\\\label{sigma}%
&&\sigma_E(x):=\sigma_E(x,x)=1+l^{-2}\delta_{ij}x^ix^j>0, \\%
\partial_P {\cal B}_l:&& \sigma_E(x)=0,
 \ee
which are invariant under the \omits{fractional linear
transformations with common denominators, }\FLT s among
Beltrami-coordinates $x^i$ 
in a transitive
form  sending the point $A(a^i)$ to the origin
$O(o^i=0)$,%
 \be\nno
x^i\rightarrow
\tilde{x}^i&=&\pm\sigma_E(a)^{1/2}\sigma_E(a,x)^{-1}(x^j-a^j)N_j^i,\\\label{FLT}
N_j^i&=&O_j^i-{ l^{-2}}%
\delta_{jk}a^k a^l
(\sigma_E(a)+\sigma_E(a)^{1/2})^{-1}O_l^i,\\\nno
O&:=&(O_j^i)_{i,j=0,\cdots,3}\in SO(4).%
 \ee 

There is an invariant for two points $A(a^i)$ and $X(x^i)$ in
${\cal B}_l$, which corresponds the cross ratio
among these two points together with the origin and  infinity in projective geometry approach: 
\be\label{AB} %
{\Delta}_{E,l}^2(a, x) = -l^2
[\sigma_E^{-1}(a)\sigma_E^{-1}(x)\sigma_E^2(a,x)-1].
\ee %

The proper length between $A$ and $B$, integral of $ds_E$ over the
geodesic segment $\overline{AB}$:
\be \label{ABL}%
L(a,b)&=& l \arcsin (|\Delta_E(a,b)|/l).
\ee

Actually, there is an important property in the Beltrami model:
the geodesics of the Beltrami metric are straight-lines  in linear
form. In fact, the geodesics can be
integrated first to get %
\be\label{q}%
{q^i}:=\sigma_E^{-1}(x)\frac{dx^i}{ds}={ consts}.%
\ee%
Thus, it is easy to see that the following rations are constants %
\be\label{um}%
\frac{q^\alpha}{q^0}=\frac{dx^\alpha}{dx^0}=consts.%
\ee%
One can integrate further to get the linear result:
\be\label{sl}%
x^i(s)=\alpha^i x^0(s)+\beta^i;\quad \alpha^i,\beta^i=consts.%
\ee%
Under the \FLT s (\ref{FLT}) of $SO(5)$, all these properties 
together with the Beltrami systems are transformed among
themselves.

In view of Klein's programm \cite{klein},  the principle of
invariance under symmetry should play a very important role in
geometry. There are also other important issues in ${\cal B}_l$
analytically.
 In view of projective geometry,  or simply the gnomonic projection,
also known as the `circle-rectilinear' transformation, the
 Beltrami-coordinates are inhomogeneous projective ones and
antipodal
identification may not be taken in order to preserve the
orientation. The great circles on (\ref{4s}) are mapped to
straight-lines, the geodesics (\ref{sl}) in ${\cal B}_l$, and vice
versa. \omits{ It is also the case for Lobachevski space \Lo$^4$
as the original model \cite{beltrami} is just for the Lobachevski
plane and completed by Klein \cite{klein1871, R}.}


In fact, the Beltrami model of momenta  with $l=a$ is just the
Euclid version of Snyder's model. What is the physics on \BdS$_l$\
in spacetime? It  just leads to the \dS-invariant
relativity in \dS-spcetime with $l=R$. 

 \subsection{The Beltarami model of \dS-space}

 \subsubsection{\dS-hyperboloid \HlsM\ and uniform `great circular' motions}

  As was noticed \cite{BdS, IWR}, via an inverse Wick rotation, the
  Riemann-sphere ${\cal S}^4$ and its Beltrami model  ${\cal B}_l$
  become the \dS-hyperboloid \HlsM\ and its Beltrami model \BdS, respectively.

  Let us first consider the \dS-hyperboloid embedded in a \Mink-space \HlsM:
\be\nno
 { H}_l:~& &\eta_{AB}\xi ^{A}\xi ^{B}\omits{-\sum
_{\alpha=1}^{3}\xi ^{\alpha}\xi ^{\alpha}-\theta\xi ^{4}\xi
^{4}}=- l^2<0, 
\\\label{qiu2} %
&&ds^2=\eta_{A B}d\xi^A d\xi^B, ~~\small
A,B=0,\cdots,4,\\\label{bdy}
\partial_P H_l: &&\eta_{AB}\xi^A\xi^B=0,%
\ee
where ${\cal J}=(\eta_{A B})=
diag (1,-1,-1,-1,-1 )$, 
 $\partial_P$ the projective boundary. They  are invariant under (linear) transformations of 
\dS-group $SO(1,4)$:
\be\label{SD14}%
\xi~ \rightarrow ~\xi'=S~\xi, \quad S {\cal J} S^t={\cal J},\quad
\forall ~S ~\in ~{SO(1,4)}.
\ee%
%

Via the \IWR, the great circles on ${\cal S}_l$ should be
`rotated' to a kind of  uniform `great circular' motions on the
\dS-hyperboloid \HlsM\ defined by a conserved 5-d
angular momentum: 
\be\label{angular5}%
  \frac{d{\cal
L}^{AB}}{ds}=0, \quad {\cal
L}^{AB}:=m_{l}(\xi^A\frac{d\xi^B}{ds}-\xi^B\frac{d\xi^A}{ds}).%
 \ee%
\omits{{ The \EoM\ for a forced particle on \dS-\HsM:}
  \be\label{angular5}\nno  %
\frac{d{\cal L}^{AB}}{ds}&=&{\cal M}^{AB},\\\nno
 {\cal
M}^{AB}&:=&\xi^A{\cal F}^B-\xi^B{\cal F}^A, \ee%
${\cal M}^{AB}$ the 5-d force-moment of a 5-d force ${\cal
F}^{A}$.\\}with an Einstein-like formula for the `mass' $m_{l}$
\begin{eqnarray}\label{5eml}%
{ -\frac{1}{2l^2}{\cal L}^{AB}{\cal L}_{AB}=m_{l}^2},\quad
{\cal L}_{AB}=\eta_{AC}\eta_{BD}{\cal L}^{CD}.
\end{eqnarray}

 There are two `time'-like scales on the \dS-hyperboloid, the coordinate-`time' $\xi^0$ and the proper-`time'$s$.
 In order to make sense for the kind of motions,
 simultaneity should be defined. As in \SR,
For a pair of two events $(P(\xi_P), Q(\xi_Q))$, they are
simultaneous in the coordinate-`time' if and only if%
\be\label{simtH}%
 \xi_P^0=\xi_Q^0.%
 \ee%
A simultaneous 3-hypersurface of $\xi^0=const$ is an expanding $S^3$：%
\be\label{s3}%
\delta_{ab}\xi^a\xi^b&=&R^2+(\xi^0)^2,~~ a, b=1,\cdots, 4;\\\nonumber
dl^2&=&\delta_{ab}d\xi^a d\xi^b. %
\ee %
For a kind of `observers' ${\cal O}_H$,\omits{ at the spacial
origin $O|\xi^\alpha=0$, where $\alpha$ takes three among
 $1,\cdots,4$, this simultaneity} it is the same with respect to the
 proper-`time'
 simultaneity in \HlsM.

 The generators of the \dS-algebra $so(1,4)$ read:
\be\label{Generator}\nno%
{i{\hat {\cal L}}}_{AB} = \xi_A \frac{\partial}{\partial\xi_B} -
\xi_B \frac{\partial}{\partial\xi_A}.
\ee%
They form an $so(1,4)$ algebraic relation of the
\dS-transformations 
on \HlsM.

The first Cisimir operator of the algebra is%
\be\label{C15}%
 \hat C_1:= -\frac 1 2 l^{-2} \mathbf{\hat {\cal
L}}_{AB} \mathbf{\hat {\cal L}}^{AB},\quad
\mathbf{\hat {\cal L}}^{AB}:=\eta^{AC}\eta^{BD} \mathbf{\hat {\cal L}}_{CD},%
\ee%
with eigenvalue   $m^2_l$, which gives rise to the classification
of the `mass'   $m_l$.

\subsubsection{The Beltrami model of \dS-space and uniform 
motions}

The Beltrami model of \dS-space (\BdS-space) is the inverse Wick
rotation of the Beltrami model of Riemann sphere.

There exist Beltrami coordinate-systems   covering \BdS-space
patch by patch. On each patch,  there are Beltrami
metric, domain condition and boundary condition  with $(\eta_{ij})_{ij=0,\cdots, 3}={diag} (1, -1,-1,-1)$%
\be\label{metric} %
{\cal B}d{\cal S}:&& ds^2=[\eta_{ij}\sigma^{-1}(x)+ l^{-2}
\eta_{il}\eta_{jk}x^l x^k
\sigma^{-2}(x)]dx^i dx^j,\\\label{domain}%
&&\sigma(x):=\sigma(x,x)=1-l^{-2} \eta_{ij}x^i x^j
>0,\\\label{bdrbds} %
\partial ({\cal B}d{\cal S}): && \sigma(x)=0. \ee%
invariant under $\cal FLT$s of $SO(1,4)$
\be\nno%
 x^i\rightarrow \tilde{x}^i&=&\pm
\sigma^{1/2}(a)\sigma^{-1}(a,x)(x^j-a^j)D_j^i,\\\label{G}
D_j^i&=&L_j^i+ { l^{-2}}%
\eta_{j l}a^l a^k (\sigma(a)+\sigma^{1/2}(a))^{-1}L_k^i,\\\nno
L&:=&(L_j^i)_{i,j=0,\cdots,3}\in SO(1,3). %
\ee

 In such a \BdS, the generators of $\cal FLT$s   read
\be\nno
  {\hat p}_i &=&(\delta_i^j-l^{-2}x_i x^j) \partial_j,~~
  x_i:=\eta_{ij}x^j,\\\label{generator}
  {\hat L}_{ij} &=& x_i {\hat p}_j - x_j {\hat p}_i
  = x_i \partial_j - x_j \partial_i \in so(1,3), 
\ee
and form an $so(1,4)$ algebra %
\be\nno%
  [ \hat{p}_i, \hat{p}_j ] &=& l^{-2} \hat{L}_{ij},~~  
  {[} \hat{L}_{ij},\hat{p}_k {]} =
    \eta_{jk} \hat{p}_i - \eta_{ik} \hat{p}_j,
\label{so14}\\
  {[} \hat{L}_{ij},\hat{L}_{kl} {]} &=&
    \eta_{jk} \hat{L}_{il}
  - \eta_{jl} \hat{L}_{ik}
  + \eta_{il} \hat{L}_{jk}
  - \eta_{ik} \hat{L}_{jl}. 
\end{eqnarray} \omits{
Thus, ${\cal SR}_{c, R}$ offers a consistent way to define a set
of observable for free particles. These issues
significantly confirm  that the motion of a free particle 
 together with the Beltrami coordinate systems and
corresponding observers  of the system are all of inertia.}

\omits{

There are two Casimir operators:
\be \label{C1}
\hat{C}_1 &=& \hat{p}_i \hat{p}^i -\frac 1 2 l^{-2} \hat{L}_{ij}
\hat{L}^{ij},\\\nno \hat{C}_2 &=& \hat{S}_i \hat{S}^i -l^{-2}
\hat{W}^2,
\ee
where
$$\hat{p}^i = \eta^{ij} \hat{p}_j,~ ~\hat{L}^{ij}=\eta^{ik}
\eta^{jl} \hat{L}_{kl},$$
$$\hat{S}_i = \frac 1 2 \eps_{ijkl} \hat{p}^j
\hat{L}^{kl}, ~ ~\hat{S}^i =\eta^{ij} \hat{S}_j, $$
$$\hat{W} =\frac 1 8 \eps_{ijkl} \hat{L}^{ij}
\hat{L}^{kl}.$$

\hfill {\small F. G\"ursey (`64);
 Lu et al (`74, `79);

 \hfill{Guo, Huang, Xu and Zhou, (`04) (`05).}

 as it was done in the relativistic quantum mechanics in
\Mink-spacetime.}}

 Note that 
for the free particles with `mass' $m_l$, the  uniform `great
circular'
 motions in the \dS-hyperboloid
having a set of conserved observables as the 5-d angular momentum
with an Einstein-like formula, now should become a kind of uniform
`motions' along straight `world'-lines in \BdS-space.

\subsection{Snyder's quantized space-time via Beltrtami model}

 It
is clear that Snyder's model can be reformulated as a \BdS-model
of momenta with $l=a$ and $ \hat{p}_i$ in (\ref{generator}) as the
spacetime coordinates' operators $\hat x_j$ in the `momentum
picture' of the quantized space-time and $\hat L^\alpha, \hat
M^\beta$ are just rest
6-generators $\hat L_{ij}$ in (\ref{so14}) of Lorentz algebra $so(1,3)$. 
Actually, the algebra (\ref{so14m})  is the same as (\ref{so14})
in the space of momenta.
Similarly, 
a quantized space-time  in  \AdS-space of momenta as an anti-Snyder's  model cam also be constructed. %
Thus, the  relation between these 
quantized space-time models and DSR (with 
$\kappa$-Poincar\'e algebra) may be described as the Beltrami
coordinates and other particular coordinate systems on 4-d
\dS/\AdS-space of momenta\cite{DSRdS}.

It is important to note that in these models after an inverse Wick
rotation the inverses of rations in (\ref{um}) become
 `group velocity' components of some
`wave-packets' and they should be also constants %
\be\label{gv}%
\frac{\partial E}{\partial p^\alpha}=consts. \quad E=p^0 c, \quad\alpha=1,2,3. %
\ee%
Thus, there is  a kind of uniform motions with component `group
velocity' or a {\it law of inertia-like} hidden in these Snyder's
models.

Since Snyder's model is a special case of the DSR, in which there
are some noncommutative aspects such as the $\kappa$-deformed
Poincar\'e algebra, there should also be some noncommutative
properties in the Beltrami model.

\subsection{Klein's Erlangen program and the principle of
relativity}

Weakening the Euclid fifth axiom leads to Riemann and Lobachevski
geometry on almost equal footing with Euclid. As was emphasized,
their physical analogies via an inverse Wick rotation are two
kinds of \dS/\AdS-invariant relativity on \dS/\AdS-spacetime,
respectively,
on the almost equal footing with Einstein's special relativity on
Minkowski spacetime. \omits{It seems that releasing the Euclid
assumption on the ruler at rest from Einstein's special relativity
leads to three kinds of
 on \dS/${
\Mink}$/\AdS-spacetime  with $SO(1,4)/ISO(1,3)/SO(2,3)$,
respectively.}

In fact, there are one-to-one correspondences between the three
kinds of geometry and their physical counterparts via the \IWR.
We list them as follows:%

\bc

 {
\begin{tabular}{lcl}
{ \quad 4-d Geometry}  &\qquad { vs.}
&\qquad {\qquad (1+3)-d Physics}\\[3mm]
${\cal S}^4/{\cal E}^4/{\cal L}^4$  &  &  \qquad
\dS$^{1,3}$/${\cal
M}^{1,3}$/\AdS$^{1,3}$\\
$SO(5)/ISO(4)/SO(1,4)$& & \qquad $SO(1,4)/ISO(1,3)/SO(2,3)$\\
Points & 　 & 　\qquad Events\\
Straight-line&  & \qquad Straight world-line \\
Principle of Invariance & & \qquad  Principle of Relativity\\
Klein  &  &\qquad  Galilei, Poincar\'e, de Sitter-Lu\\
Erlangen Program\qquad  &  &  \qquad Theories of Relativity\\
$\cdots$ \qquad  &  &  \qquad   $\cdots$\\

\end{tabular}}

 \ec

It should be noted that the 4-d Riemann-sphere ${\cal S}^4$ may be
regarded as an instanton with an Euler number $e=2$ in the sense
that it is a solution of the Euclidean version of gravitational
field equations, its  quantum tunnelling scenario should support
$\Lambda>0$ as the \BdS\ \cite{IWR}. It will be shown that in the
simple model of the \dS-gravity \cite{dSG, T77, QG} this is the
case as in the \GR.

%
%

\section{De Sitter Invariant  Relativity}

\subsection{Inertial motions, transformations and the principle of relativity}


In general, we may define the inertial motions as a kind of the
uniform (coordinate) velocity  motions along a straight line.
Namely, in a (coordinate) system $S(x)$ the motions satisfy
\be\label{uvm}%
 x^\alpha=x_0^\alpha+v^\alpha(t-t_0),\quad
v^\alpha=\frac{dx^\alpha}{dt}=consts,\quad \alpha=1, 2, 3, %
\ee
the motions are called the inertial ones and the system the
inertial system ($\cal IS$). These are the same as in Newton's
mechanics and Einstein's special \SR.  The differences among them
are that 
the proper length of a
 rigid ruler or the the proper time of a ideal clock are not assumed to obey the Euclid geometry. In other words,
 the spacial coordinates themselves and the temporal coordinate itself are not assumed to be uniform.   This
 leads to  the transformations among $\cal IS$s are different.

\omits{Thus, we may start with eqn. and to get \omits{what Fock
proved to get Fock's theorem as a lemma denoted as }the lemma 1.}
\omits{Fock proved a theorem on the general transformations among
uniform velocity motions \cite{Fock}.}

In an inertial (coordinate) system $S$, there is a particle with
uniform velocity motion (\ref{uvm}). If in the transformed $\cal
IS$  $S'$, the same particle is described by
\be\label{uvm'}%
{x'}^\alpha={x'}_0^\alpha+{v'}^\alpha(t'-t'_0), \quad
 {v'}^\alpha=\frac{d{x'}^\alpha}{dt'}=consts,
 \ee
what are the most general transformations? Due to Umov, Weyl and
Fock \cite{Fock}, the most general form of the transformations should be %
\be\label{FockT}%
{x'}^i=f^i(t, x^\alpha),\quad i=0, \cdots, 3, %
\ee
which transform a uniform straightline motion in $S$ with
(\ref{uvm}) to a motion of the same nature in $S'$ with
(\ref{uvm'}) are that the four functions $f^i$ are ratios of
linear functions, all with the same denominators, i.e. the  $\cal
FLT$s.

Further, we should require that there exist a  metric in 4-d
spacetime, in which there are inertial systems,
 and the $\cal FLT$s form a group with ten parameters, like the Galilei
 group in Newton's mechanics and Poincar\'e group in Einstein's
 special \SR, including four for spacetime `translations', three for boosts,  and the rest three for space
 rotations.
 Thus, according to the properties of maximally symmetric spaces \cite{KN}: such kind of 4-d spaces
 with metric invariant under
ten-parameter transformation groups should be maximally symmetric
spaces
of constant curvature with 
radius $R$ or $R \rightarrow \infty$. Namely, they are the
 \dS/\Mink/\AdS~space with  $SO(1,4)/ISO(1,3)/SO(2,3)$-invariance,
respectively.


Thus, for the \dS/\AdS-spacetime, the Beltrami systems are just these systems. 
Therefore, on the \BdS/anti-\BdS-spacetime, there  are  also the
principle of relativity and the postulate on universal constants.
The principle
 requires that {\it all physical laws without gravity are invariant
under the \FLT s of \dS/\AdS-group among the inertial systems,
respectively.} The postulate  states that {\it in $\cal IS$s on
4-d spacetimes, there are two universal constants:  the speed of
light $c$ and the
length $R$ as 
the curvature radius.}

As was mentioned for the Beltrami model of Riemann sphere, one
patch of the Beltrami coordinate cannot cover the whole
\dS/\AdS-spacetimes, but the later can be covered by the Beltrami
coordinate
systems patch by patch and all 
transition functions on intersections between different patches
are of \FLT-type \cite{BdS}.

In each patch, say,
$U_{4}, \xi^4 > 0$,  the Beltrami coordinates are%
 \be \label{u4}%
x^i|_{U_{4}} =R \frac{\xi^i}{\xi^4},\quad i=0,\cdots, 3;\quad
\xi^4=({\xi ^0}^2-\sum _{a=1}^{3}{\xi ^a}^2+ R^2 )^{1/2} > 0,
\ee%
there are Beltrami metric (\ref{metric}), 
condition (\ref{domain}) and  boundary (\ref{bdrbds}) with $l=R$ \omits{ %
\be\label{bhl}%
 ds^2&=&[\eta_{ij}\sigma^{-1}(x)+ R^{-2}
\eta_{il}\eta_{jk}x^l x^k \sigma^{-2}(x)]dx^i dx^j,
\\\label{domain}%
&&\sigma(x)=\sigma(x,x):=1-R^{-2} \eta_{ij}x^i x^j
>0, \\\label{bdy2}%
\partial_P{\cal B}d{\cal S}: &&\sigma(x)=0. %
 \ee%
}invariant under $\cal FLT$s (\ref{G}) 
of $SO(1,4)$ with $l=R$, \omits{
\be\label{G}\nno
x^i\rightarrow &&\tilde{x}^i=
\sigma^{1/2}(a)\sigma^{-1}(a,x)(x^j-a^j)D_j^i,\\
&&D_j^i=L_j^i+ { R^{-2}}%
\eta_{j l}a^l a^k (\sigma(a)+\sigma^{1/2}(a))^{-1}L_k^i,\\\nno
&&L:=(L_j^i)_{i,j=0,\cdots,3}\in SO(1,3), %
\ee%
}which transform a point $A(a), \sigma(a)>0 \in$\BdS\ to the origin. \omits{, $\eta_{ij}%
={diag} (1, -1,-1,-1)$, }
It is clear that the $\cal IS$ $S(x)$ 
maps to  $\cal IS$ $\tilde S(\tilde x^i)$ and 
the inertial motions in $S$ are transformed to that in $\tilde S$.
Namely, the geodesics are straight world-lines in linear forms,
and vice versa, and transformed among themselves.

Thus,  there is the law of inertia in \BdS/anti-\BdS: {\it The
free particles and light signals without undergoing any unbalanced
forces should keep their uniform motions along straight
world-lines in the linear forms in \BdS/anti-\BdS-space,
respectively.}

The equation  of motion for a forced particle can also be given
\cite{BdS2}.

 For free  particles there is a set of inertial conserved
quantities $p^i, ~L^{ij}$ along a geodesic, \omits{}
\be\label{angular4}%
 p^i=\sigma(x)^{-1}m_{R}\frac{d x^i}{ds},&& \frac{dp^i}{ds}=0;
 \\\nno%
 L^{ij}=x^ip^j-x^jp^i, &&\quad\frac{dL^{ij}}{ds}=0.
\ee%
These are 
the pseudo 4-momentum $p^i$, pseudo 4-angular-momentum $L^{ij}$ of
the particle. They constitute a conserved 5-d angular momentum as
was shown in
(\ref{angular5}) with $l=R$%
\omits{ \be\label
 \frac{d{\cal L}^{AB}}{ds}=0,\quad  {\cal
L}^{AB}:=m_{R}(\xi^A\frac{d\xi^B}{ds}-\xi^B\frac{d\xi^A}{ds}). %
\ee%
T} and  satisfy a generalized Einstein  formula  in \BdS\ from the Einstein-like formula (\ref{5eml}): %
\begin{eqnarray}\label{eml}%
 E^2=m_{R}^2c^4+{p}^2c^2 + \d
{c^2} {R^2} { j}^2 - \d {c^4}{R^2}{
k}^2,%
\end{eqnarray}
with energy $E=p^0$, momentum $p^\alpha$,
$p_\alpha=\delta_{\alpha\beta}p^\beta$, `boost' $k^\alpha$,
$k_\alpha=\delta_{\alpha\beta}k^\beta$ and 3-angular momentum
$j^\alpha$, $j_\alpha=\delta_{\alpha\beta} j^\beta$. Note that
$m^2_{R}$ now is the eigenvalue of 1st Casimir operator of
\dS-group, the same as the one in (\ref{C15}) with $l=R$.

If we introduce the Newton-Hooke constant $\nu$
\cite{NH} and link the curvature radius $R$ with the cosmological constant $R\simeq (3/\La)^{1/2}$%
\be\label{NHc}%
\nu:=\d {c} {R}\simeq c (3/\La)^{-1/2},\quad \nu^2 \sim 10^{-35}s^{-2}. %
\ee%
It is very tiny. Thus, local experiments on ordinary scales can
not distinguish  the \dS-invariant relativity from Einstein's
special relativity.

The interval between two events and  thus the light-cone can be well
defined as the inverse Wick rotation counterparts of (\ref{AB}) and
(\ref{ABL}), respectively.

In fact, for two separate events $A(a^i)$ and $X(x^i)$ in \BdS,
\be\label{lcone0} %
 \Delta_R(a, x)^2 = R^2\,[\sigma(a)^{-1}\sigma(x)^{-1}\sigma(a,x)^2-1]%
\ee %
is invariant under the \FLT s of $SO(1,4)$, 
 Thus, the interval between $A$ and
$B$ is timelike, null or spacelike,
respectively, according to%
\begin{equation}\label{lcone}%
\Delta_R^2(a, b)\gtreqless 0.%
\end{equation}
The proper length of time/space-like 
interval between $A$ and $B$ is the integral of $ds$ over the
geodesic segment $\overline{AB}$:
\be \label{AB1}%
S_{t-like}(a, b)&=&R \sinh^{-1} (|\Delta(a,b)|/R), \\
\label{AB1sl} S_{s-like}(a,b)&=& R \arcsin (|\Delta(a,b)|/R).%
\ee
\omits{where ${\cal I}=1, -i$ for timelike or spacelike,
respectively.}

The light-cone at $A$ with running points $X$ is
\be \label{nullcone} %
{\cal F}_{R}:= R
\{\sigma(a,x) \mp [\sigma(a)\sigma(x)]^{1/2}\}=0.%
 \ee%
It satisfies the null-hypersurface condition.\omits{
 \be\label{Heqn}%
\left . g^{ij}\frac{\partial f}{\partial x^i}\frac{\partial
f}{\partial x^j}\right |_{f=0}=0, %
\ee
where $g^{ij}=\sigma(x)(\eta^{ij}-R^{-2} x^i x^j)$ inverse of
(\ref{bhl}).} %
At the origin %
$a^i=0$, the light cone becomes a Minkowski one and 
$c$ is numerically the velocity of light in the vacuum.

There is a horizon tangent to the boundary in \BdS\ for the
observers ${\cal O}_I$:
\be\label{Heq}%
\lim_{a \to a'} \sigma (a,x)=0, \qquad \lim_{a\to a'}\sigma(a)=0.%
\ee

\subsection{Two kinds of  simultaneity, the  principle of relativity and cosmological principle}

In order to make physical measurements, one should  define
simultaneity. Different from Einstein's special relativity, there
are two kinds of simultaneity related to two kinds of
measurements, or to the principle of relativity and the
cosmological principle, respectively, in \dS/\AdS-spacetime. In
the contraction $R\to \infty$, they coincide with each other.

\subsubsection{The Beltrami-time simultaneity}

 Let us first consider the Beltrami coordinate simultaneity, called the 
 Beltrami simultaneity.
 For inertial observers ${\cal O}_I$ at spacial origin, two events ($A, B$) are simultaneous if and only if  
\be %
a^0:=x^0(A) =x^0(B)=:b^0. 
\ee%
It defines a $1+3$ decomposition of \BdS-space
\be%
 ds^2 =  N^2 (dx^0)^2 - h_{ab} \left (dx^a+N^a dx^0 \right
)
\left (dx^b+N^b dx^0 \right ) %
\ee
with lapse function, shift vector and induced 3-geometry on
3-hypersurface $\Si_c$ in one coordinate patch, respectively
%
\begin{eqnarray}\nno
N\ &=&\{\si_{\Si_c}(x)[1-(x^0 /R)^2]\}^{-1/2}, \\%
N^a&=&x^0 x^a[ R^2-(x^0)^2]^{-1},  \\\nno h_{ab}&=&\dl_{ab}
\si_{\Si_c}^{-1}(x)-{ [R\si_{\Si_c}(x)]^{-2} \dl_{ac} \dl_{bd}}x^c
x^d ,\\\nonumber%
 &&\si_{\Si_c}(x)=1-(x^0{/R})^2 + {\dl_{ab}x^a x^b
/R^2}.\nno
\end{eqnarray}
%

%

It is easy to see that at $x^0=0$, $\si_{\Si_c}(x)=1+{\dl_{ab} x^a
x^b/R^2}, ~N=\si_{\Si_c}^{-1/2}(x), ~N^a=0.$ Then the Cauchy
hypersurface is $ \Si_c \simeq S^3$.
 And at  Beltrami time $x^0\neq 0$, as long as $x^0$ is still time-like, we should also have {$ \Si_c \simeq S^3$}.

The Beltrami simultaneity leads \omits{is easily generalized} to
the definition of Beltrami ruler and its relation to the spacial distance of two events. A Beltrami non-Euclidean ruler
at time $x^0$ is given by 
\be%
dl_B^2|_{x^0}=-h_{ab}|_{x^0}dx^a dx^b.%
\ee

\subsubsection{The Proper-time simultaneity  and the Robertson-Walker-like 
 coordinates}

The proper time $\tau$ of a rest clock on the time axis of
Beltrami
 system, $\{x^a=0\}$, reads
\begin{eqnarray}\label{ptime}%
 \tau=R \sinh^{-1} (R^{-1}\sigma^{-\frac{1}{2}}(x)x^0).
\end{eqnarray}
one can naturally define the second simultaneity by means of
$\tau$ as 
the proper-time simultaneity. The events are simultaneous with
respect to the proper time of a clock rest at the origin of the
Beltrami spatial coordinates if and only if these events
corresponding to the same $\tau$
\begin{equation}\label{smlt}
 x^0\sigma^{-1/2}(x,x)=(\xi^0:=)R \sinh(\tau/R)=\rm const.
\end{equation}
 If $\tau$ is taken as a `cosmic' temporal coordinate together with the spatial Beltrami coordinates,
 the Beltrami system becomes
a Robertson-Walker-like system with a metric:%
\be\nno%
 ds^2&=&d\tau^2-dl_C^2=d\tau^2- \cosh^2(\tau/R) dl_{0}^2,\\\label{dsRW}%
dl_{0}^2&=& {\{\delta_{ab}\sigma_{\Sigma_\tau}^{-1}(x)
-[R\sigma_{\Sigma_\tau}(x)]^{-2}\delta_{ac}\delta_{bd}x^c x^d\}}
 dx^a dx^b, \\\nno 
&& \sigma_{\Sigma_\tau}(x,x)=1+R^{-2}\delta_{ab}x^a x^b
>0,
\ee
%
with $dl_{\omits{\Sigma_\tau} 0}^2$ is a 3-dimensional
Beltrami-metric on an $S^3$ with radius $R$. It is  an `empty'
accelerated expanding cosmological model with a slightly closed
cosmos in  $O(R)$.

There is an important prediction different from the `standard model' 
if the dark universe is asymptotically a \dS, the 3-d cosmic space
is asymptotically an expanding $S^3$
in Robertson-Walker-\dS\ spacetime. If we take take $R^2\simeq 3\Lambda^{-1}$,  
the deviation from the flatness is in $O(\Lambda)$.
\omits{It is already indicated by the data from WMAP: %
The $\Omega_T =1.02\pm 0.02$ and angular power
spectrum\cite{WMAP}. Of course, it should be further checked.

In addition, there is a preferred time direction in \dS\ picked up
by the evolution of the dark universe.}

Due to the relation between the principle of relativity and
cosmological principle, \dS-space provides also such a model that
the \dS-cosmic background with cosmological constant just acts as
{\it the origin of the inertial law }\omits{, i.e. the origin of
inertial motions and inertial coordinates, {\it the origin of
inertia} for short,}
 This supports as a base the principle of relativity in
Beltrami-coordinates. Of course, the precondition is that the
maximum symmetry ensures the existence of the inertial motions and
these two principles in \dS-spacetime. In other words,  the
\dS-group as a maximum symmetry assures that there are the \PoR\ and
\CP\ together with their relation in the \dS-spacetime as a
maximally symmetric one. Thus, the Robertson-Walker-\dS-cosmos with
cosmological constant and  other cosmic objects including the
distant stars  as test stuffs should just display as the origin of
\dS-inertial law in the Beltrami-coordinates on \dS-spacetime.

In fact,  in \dS-space there are a type of {\it inertial-comoving
observers} having a kind of two-time-scale timers 
with respect to the Beltrami-time and the
cosmic-time. This reflects that there is an important relation
linking the principle of relativity with the cosmological
principle of \dS-invariance.

\bc  \begin{tabular}{lcl} {  Inertial  ${\cal O}_I$}  \quad &{
vs.}
&\quad { Co-moving  ${\cal O}_C$}\\
Beltrami-systems \quad & 　&\quad Robertson-Walker-\dS-systems\\
Beltrami timer \quad & 　&\quad  Cosmic-time timer\\
Beltrami ruler \quad & 　&\quad  Co-moving ruler\\
Inertial observables \quad & 　&\quad Co-moving observables\\
$\qquad  \cdots $ \quad & 　&\quad  $\qquad\cdots$
\end{tabular}
\ec

Thus, what should be done for those inertial-comoving observers is
merely to switch their timers from cosmic-time back to
Beltrami-time 
and vice versa. Namely, once the observers would carry on the
experiments in their laboratories, they should take their timers
switching on Beltrami-time  and off the cosmic-time  so as to act
as inertial observers and all observations are of inertial. When
they would  take cosmic-observations on the distant stars and the
cosmic objects other than the cosmological constant as test stuffs
they may just switch off the Beltrami-time and on the cosmic-time
again, then they should act as a kind of comoving observers as
they hope.

\omits{In fact, the realizations of \dS-symmetry is the base of
the \dS-cosmos with cosmological constant as the origin of
\dS-inertial law in the Beltrami-coordinates of the
\BdS-spacetime.}

\omits{{\it \dS-cosmos with cosmological constant as origin of
inertial law.} Due to the relation between \PoR\ and \CP,
\dS-space provides also such a model that {\it the origin of the
inertial law }\omits{, i.e. the origin of inertial motions and
inertial coordinates, {\it the origin of inertia} for short,} is
just the cosmos with cosmological constant. This supports as a
base the principle of relativity in Beltrami-coordinates.

In fact,  on \dS-space there is a type of {\it inertial-comoving
observers} having a kind of two-time-scale timers of the
Beltrami-time and the cosmic-time with \omits{the two kinds of
simultaneity with respect to }an important relation between the
principle of relativity and cosmological principle. What should be
done for those inertial-comoving observers is merely to switch
their timers from cosmic-time back to Beltrami-time according to
their relation and vice versa. Namely, once the observers would
carry on the experiments in their laboratories, they should take
their timers switching on Beltrami-time scale and off the
cosmic-time scale so as to act as inertial observers and all
observations are of inertial while when they would  take
cosmic-observations on the distant stars and so on as test
particles they may just switch off the Beltrami-time scale and on
the cosmic-time scale again, then they should act as a kind of
comoving observers as they hope.}

\subsection{On temperature and entropy}

In \GR,  the Hawking-temperature and entropy at the \dS-horizon
\cite{GH77} lead to the \dS-entropy puzzle. \omits{: Why {is a}
\dS-spacetime of constant curvature like a black hole and what is
the statistical origin of the entropy?}

There is another 
explanation now \cite{T}. Eq. (\ref{ptime}) shows  the imaginary
Beltrami-time\omits{ and the imaginary proper-time,} has no
periodicity,\omits{ for the former} since both Beltrami-time axis
and its imaginary counterpart are
straight-lines without coordinate acceleration, but the imaginary
proper-time has  such a period  that  is inversely proportional to
the Hawking-temperature $(2\pi R)^{-1}$ at the horizon. If the
temperature Green's function can still work, this should indicate
that the horizon in Beltrami-coordinates is at zero-temperature and
needless to introduce entropy. \omits{In addition, it can be shown
that the so-called `surface-gravity' at the horizon in, say, the
static \dS-coordinates \omits{the Robertson-Walker-like ones} is
just the coordinate acceleration caused by non-inertial motions in
view of the inertial observers. \blue [There is an important
difference between de Sitter spacetime and Rindler spacetime.  In
Rindler spacetime, the observer itself experiences an acclerlation.
So, he moves non-inertially. While, in the de Sitter spacetime, the
observer moves inertially (in the sense of trajectory) similar to
the observer at infinity in Schwarzschild spacetime. Instead, the
`static' test body on the horizon experiences an acceleration. (The
situation for Schwarzschild black hole shares the same property.)
Therefore, simply to say the coordinate acceleration is caused by
non-inertial motions seems to be questionable.  (It is this reason
that I added `and non-inertial parameterization' in the end of the
present paragraph.) ]}

Since there is no gravity, the Hawking-temperature and
(area-)entropy $S$ at the horizon in other coordinate-systems
 should be non-inertial effects \cite{T}. Such a kind of
non-inertial thermodynamic properties for \dS-spacetime are similar
to those in Rindler metric in flat spacetime.

Similar issues may also be introduced and make sense in the
\dS-space of momenta.

\omits{
\subsection{\dS-cosmos with $\La$ as origin of inertial law}

%
%

\omits{The origin of the law of inertial motion (\OoI) is a long
standing open problem.\\}

Mach: { ``When, accordingly, we say, that a body preserves
unchanged its direction and velocity in space, our assertion is
nothing more or less than an abbreviated reference to { the entire
universe}." ``How can we determine such a frame? Only by referring
to other bodies in the universe."}


\omits{
Einstein:\\

 Mach's principle (1918).\vskip 5mm

{ ``The G-field is without remainder determined by the masses of
bodies. Since mass and energy are, according to results of the
special theory of relativity, the same, and since energy is
formally described by the symmetric energy tensor ($T_{\mu\nu}$),
this therefore entails that the G-field be conditioned and
determined by the energy tensor. "}\\

It gets rid of the action-at-a-distance!\vskip 5mm

But mixes up the \OoI~ with the \OoIm, and the free fall motions,
the local inertial motions and  inertial systems.\\

Experiments fail to prove it!

\hfill{\small (See, e.g. S. Weinberg 1972.)}\\

Einstein gave up his Mach's principle in 1954. }

{ Remarks:} Mach could make sense only if 1. there might be a
 cosmos picture theoretically;
 2. The picture should be in consistent with the principle of
 relativity!
 Then, in the sense of Mach: in such a theory, the cosmos should determine the law of inertia and the inertial
 frames.
 Otherwise, Mach's statements are nonsense!

\omits{{ - To distinguish the origin of inertial mass from the
origin of the law of inertia, and of the local inertial motion,
i.e.
the  geodesics motions in \GR.}\vskip 3mm

  In the spirit of Riemann, Mach, and Einstein, based on recent
observations on dark sector with cosmological
constant, Mach's statement should be revised.}

Mach's principle in dark universe: {\it Both the  \OoI~ and the
one of local inertial motion should be dominatingly determined by
the {dark sector with $\La$} along with a little from the distant
stars and others.} There is an important  logic consequence: {\it
in \dS-space without any `matter' except the distant stars as the
test `particles', the inertial motions should exist, and the \OoI\
should be the \dS-cosmic background with  $\La$}.

{\BdS-spacetime and its Robertson-Walker-like counterpart just
provide such a model:} the \dS-cosmos with  $\La$ is the \OoI\ in \BdS.

A set of inertial-comoving observers ${\cal O}_{I-C}$:
 Inertial experiments  $\leftrightarrow$ timer with respect to
Beltrami-time. Co-moving observations $\leftrightarrow$ timer with
respect to cosmic-time. Inertial observers ${\cal O}_{I}$, 
with respect to inertial observable. 
 Cosmic observers ${\cal O}_{C}$, with respect to co-moving
observable. From one to another:
Referencing 
the Beltrami-time $x^0$ but not the cosmic-time $\tau$, and vice
versa!

 If $\La \to 0$, ${\cal SR}_{c, \Lambda\gtrless 0}$ tend to
${\cal SR}_{c}$, i.e. ${\cal SR}_{c, \Lambda \to 0}$: Two kinds of
simultaneity coincide. The set of { \PoR-$\cal MC$} puzzles
reappear, and \OoI~opens again.

\omits{What  would you prefer?

{\it Keep on $\cal GR$ with respect to the set of $\cal SRP-MPC$
puzzles ?!

Take an adventure on $dS$ with respect to $\cdots$ ?!}


 This is, in fact, a prediction:\\
  The 3-d cosm'l space is closed if $\Lambda > 0$ , the
deviation from flatness is order of $\Lambda$ !}

}

%
%

\section{The
Planck Scale - $\Lambda$ Duality }

\subsection{An interchangeable relation and the duality}

It is interesting to see that  there is an interchangeable dual
relation between Snyder's model and \dS-invariant \SR\ in  \BdS\
with  $l$:
 \bc
\begin{tabular}{lcl}
{ \quad Snyder's QST}  \quad &  &\quad
 {\dS-invariant \SR}\\
momentum `picture' \quad &  &\quad coordinate `picture'\\
\BdS-space of momenta \quad &  &\quad  　\BdS-spacetime\\
$l=a \sim$ Planck length\quad &  &\quad  $l=R \sim$ cosmic  radius \\
constant `group velocity'\quad &  &\quad  constant 3-velocity\\
quantized space-time \quad &  &\quad  `quantized' momenta \\
$\qquad\hat x_\alpha, \hat t$\quad &  &\quad  $\qquad\hat p_\alpha, \hat E$\\
$ \tilde{T}_p=0$, without $\tilde{S}_p$  \quad &  &\quad $T=0$,
without $S$\\
\end{tabular}
\ec

It is important that  from two fundamental constants, the Planck
length $l_P:=( G\hbar c^{-3})^{1/2}$ and the \dS-radius
$R\simeq(3/\La)^{1/2}$, it follows a dimensionless constant 
\be%
 {\iota}:=\kappa/R \to g:=\sqrt{3}\ell_P/R, ~~ g^2\simeq G\hbar
c^{-3}\Lambda\sim 10^{-122}.%
\ee%
Since there is Newton constant, $g$ should describe the gravity
and its self-interaction with local \dS-invariance between these
two scales.


Thus, these indicate that there should be a Planck
scale-cosmological constant duality: {\it The cosmological
constant is a  fundamental scale as the Planck length.
 The physics at
such two  scales should be dual to each other in some `phase' space
and linked via the gravity with local \dS-invariance characterized
by the dimensionless constant $g$.}


 It is clear that the Planck scale-$\Lambda$ duality could also be proposed for the
\AdS-space. And  this is a kind of `ultraviolet-infrared' relations
since the cosmological constant and the Planck length provide an IR
and a UV cut-off, respectively.
 The `UV-IR duality' (or connection, etc) appeared in various
cases already. For example, in the sense of \AdS/$CFT$
correspondence and holographic principle \cite{sussk, Peet}. The
relations among these UV-IP should be investigated.

\subsection{Is the dimensionless constant calculable?}
It is interesting to see that $ g^2$ is in the same order of
difference between $\Lambda$ and the theoretical quantum `vacuum
energy', there is no longer this puzzle in view of the
\dS-invariant relativity and gravity with local \dS-invariance.
However, since $\Lambda$ a fundamental constant as  $c, G$ and
$\hbar$, a further question should be: What is the origin of the
dimensionless
 constant $g$?
 Is it calculable? This is just  the first question of the `top ten' \cite{top10}: `Are
all the (measurable) dimensionless parameters that characterize
the physical universe calculable
 in principle or are some merely determined by historical or quantum mechanical accident and
 uncalculable?'

It is important to note that there are some hints on the answer
for this dimensionless constant $g$. First,  among 4-d Euclid,
Riemann and Lobachevski spaces there is only the Riemann-sphere
with non-vanishing 4-d topological number. Thus, the  quantum
tunneling scenario for the Riemann-sphere ${\cal S}^4$ as an
instanton of gravity to the \dS-space may explain why the
cosmological constant should be positive, i.e. $\Lambda>0$.  We
should show in the next subsection that in a simple  model of
\dS-gravity in \dS-Lorentz gauge shows that this is just the case.
Further, if the action of the \dS-gravity is of the Yang-Mills
type, then its Euclidean version is of a non-Abelian type with
local $so(5)$ symmetry. Thus, due to  asymptotic freedom, the
gauge-coupling constant, say $g$,
 should be running and
approaching to zero as the momentum reaches to infinity. However,
for the case of gravity, the momentum could not be reaching to
infinity but the Planck scale as a fixed point so that the Euclidean
counterpart of the dimensionless coupling constant should be very
tiny and link $\Lambda$ with the Planck scale.


\subsection{A simple model of \dS-gravity}

In \GR, there is no room for special relativity in \dS/\AdS-space.
In \dS/\AdS-invariant relativity,  there is no gravity in
\dS/\AdS-space.  How to describe gravity?

As the spirit of Einstein's equivalence principle, the gravity
should be based on localized  special \SR. However, in general
relativity there are only
 local Lorentz-frames of homogeneous Lorentz-symmetry without localized translations.

\omits{There is  a \ul{`Gordian knot'} i}In Einstein equation
${\bf G}=8\pi G{\bf T}$ symbolically (see, e.g.  \cite{MTW}), the
Einstein-Cartan `moment of rotation' $\bf G$ is related  to  local
homogeneous Lorentz rotation, while the stress-energy tensor $\bf
T$ is concerning the translations in \Mink-space (see, e.g.
\cite{trautm}). Why geometry is connected with matter in different
symmetry?

As an enhanced equivalence principle with  localization of special
\SR, there should be  {\it the localization principle}: {\it On
spacetimes with gravity, there always exist local
relativity-frames with  local Poincar\'e/dS/\AdS-symmetry,
physical laws   must take the gauge covariant versions of their
special-relativistic forms with respect to the local
Poincar\'e/dS/\AdS-symmetry, respectively.} If geometry and matter
are connected in same local symmetry, \omits{n order to avoid the
\ul{`Gordian knot'}, }there
 should be some
gauge-like dynamics for gravity. 

A simple model for the \dS-gravity  has shown such a feature. Its
action is  gauge-like and  characterized by a dimensionless
constant $g$. In the \dS-Lorentz gauge, it reads \cite{dSG, T77}
\be\nno%
 S_G&=& -\d {1}{4g^2} \int_{\cal M} d^4 x \varepsilon ({\cal
F}^{AB}_{~~~jk} {\cal F}_{AB}^{~~\,jk})\\%
&=&\int_{\cal M} d^4x \varepsilon \left (\d 1 {16\pi G} (F-2\La)
-\d {1}{4g^2} F^{ab}_{~~jk} F_{ab}^{~~jk} + \frac 1 {32\pi G}
T^{a}_{~jk}T_a^{~jk}
\right ),%
\ee
%
where $\varepsilon=det(e^a_{~j})$, ${\cal F}^{AB}_{~~~jk}$ is the
\dS-curvature of a \dS-connection ${\cal B}^{AB}_{~~j}\in
so(1,4)$, with ${\cal B}^{ab}_{~~j}=B^{ab}_{~~j}, {\cal
B}^{a4}_{~j}=R^{-1}e^a_{~j}$, $F$, $F^{ab}_{~jk}$ and
$T^{a}_{~jk}$
 are scalar curvature, curvature and torsion of the
Riemann-Cartan manifolds ${\cal M}$ with Lorentz frame $e^a_{~j}$
and connection $B^{ab}_{~j}$. Namely,
\be\label{dSLA}%
{\cal B}:={\cal B}_jdx^j, ~ {\cal B}_j:=( {\cal
B}^{AB}_{~~j})_{A,B=0,\cdots, 4}= \left(
\begin{array}{cc}
B^{ab}_{~~j} & R^{-1} e^a_j\\
-R^{-1}e^b_{j} &0
\end{array}
\right ) \in {so}(1,4), %
\ee%
where $R$ is the \dS-curvature radius. The curvature valued {in}
the \dS-algebra reads:
\be\label{dSLF}%
{\cal F}_{jk}= ( {\cal F}^{AB}_{~~jk})&=&\left(
\begin{array}{cc}
F^{ab}_{~~jk} + R^{-2}e^{ab}_{~~ jk} & R^{-1} T^a_{~jk}\\
-R^{-1}T^b_{~jk} &0
\end{array}
\right ) \in { so}(1,4),
\ee%
where $e^a_{~bjk}=e^a_je_{bk}-e^a_ke_{bj}, e_{bj}=\eta_{ab}e^a_j$,
$ F^{ab}_{~~ jk}$ and $ T^a_{~jk}$ are given by 
\be\omits{\label{T2form}%
\Omega^a&=&d\theta^a+\theta^a_{~b} \wedge
\theta^b=\frac{1}{2}T^a_{~jk}dx^j\wedge dx^k\\\label{F2form}
\Omega^a_{~b}&=&d\theta^a_{~b}+\theta^a_{~c}\wedge\theta^c_{~b}=\frac{1}{2}F^a_{~b
jk}dx^j\wedge dx^k;\\}\label{Ta}
T^a_{~jk}&=&\partial_je^a_k-\partial_ke^a_j+B^a_{~c
j}e^c_k-B^a_{~c
k}e^c_j,\\\label{Fab}%
F^a_{~b jk}&=&\partial_jB^a_{~bk} -\partial_kB^a_{~bj}+B^a_{~cj}B^c_{~bk}-B^a_{~ck}B^c_{~bj}.%
\ee%
where $B^a_{~bj}=\eta_{ac}B^{ac}_{~~j}$. And
$F=\frac{1}{2}F^{ab}_{~jk}e_{ab}^{~jk}$.

 It is important that this model can pass all tests for \GR\ and may provide a more suitable platform for the precise
cosmology.

\omits{Remarkably, $g^2$ is in the same order of difference
between $\Lambda$ and the theoretical quantum `vacuum energy', the
big difference is no longer  a puzzle  in \dS-invariant relativity
and local \dS-invariant gravity. Since $\Lambda$ is  fundamental
as $c$, $G$ and $\hbar$,  further questions should be: What are
the origins of these fundamental constants or the origin of the
dimensionless constant $g$? Are they calculable?

For 4-d Euclid/Riemann/Lobachevsky-space, just the Riemann-sphere is
an instanton with non-trivial topology. Its quantum tunnelling
scenario supports $\Lambda>0$. Since the Euclidean 
action is of  $so(5)$-gauge type, the asymptotically freedom
mechanism may indicate the gauge-coupling constant should be running
to zero as the momentum to infinity. For the gravity, the momentum
could only be approaching the Planck scale $\ell_P=(G\hbar
 /c^{3})^{1/2}$ as a fixed point, so $g$ should be very tiny and link $\Lambda$ with
 $\ell_P$.

}

\section{Concluding  Remarks}
With plenty of \dS-puzzles, the dark universe  as an accelerated
expanding, asymptotic to \dS-space with a tiny cosmological
constant $\Lambda$ \cite{Riess98, WMAP}  greatly challenges
Einstein's theory of relativity as foundation of the physics in
large scale.

Symmetry, its localization and symmetry breaking play extreme
important roles in physics. For the space-time and gravity
physics, the maximum symmetry and its localization should also
play an extreme important role. There should be three kinds of
relativity and their contractions \cite{NH}, and
 three kinds of theory of gravity as localization of corresponding relativity with a gauge-like dynamics
 and their contractions, respectively. Our Nature should chose one
 of them.

Via an interchangeable dual relation between Snyder's-like quantized
space-time models in \dS/\AdS-space of momenta and the
\dS/\AdS-invariant special relativity in \dS/\AdS-spacetime with the
Beltrami coordinates, respectively, there should also be  a duality
in the physics at the Planck scale and cosmological constant. And
between these two scales is the gravity based on the localization of
corresponding \SR\ with a gauge-like dynamics of full localized
symmetry characterized by a dimensionless coupling constant.

{ The dark universe may already indicate that the \dS-invariant
relativity and the gravity with local
 \dS-invariance should be the foundation of physics in large-scale.}

Needless to say, there is still long way to go!

\begin{acknowledgments}
The author would like to thank Professor/Dr. Q.K. Lu, Z. Chang,
C.-G. Huang, Y. Tian,  Z. Xu and B. Zhou for valuable discussions.
This work is partly supported by NSFC under Grant Nos 10375087,
90503002.
\end{acknowledgments}


\end{document}